\documentclass[a4paper,10pt]{article}
\usepackage[latin1]{inputenc}
\usepackage[english]{babel}
\usepackage{amsmath,amsthm}
\usepackage{amssymb}
\usepackage{ifpdf}
\ifpdf
  \usepackage[pdftex]{graphicx}
  \usepackage{epstopdf}
\else
  \usepackage[dvips]{graphicx}
\fi
\usepackage[margin=3.5cm]{geometry}
\usepackage[pdftitle={URDME manual},
 pdfauthor={Engblom},
 pdffitwindow=true,
 breaklinks=true,
 colorlinks=true,
 urlcolor=blue,
 linkcolor=red,
 citecolor=red,
 anchorcolor=red]{hyperref}
\usepackage[sort&compress,numbers]{natbib}
\usepackage{algorithm}
\usepackage{algorithmic}

\numberwithin{equation}{section}
\numberwithin{table}{section}
\numberwithin{figure}{section}

\newcommand{\urdme}{\texttt{urdme}}
\newcommand{\umod}{\texttt{umod}}
\newcommand{\mex}{\texttt{mex}}

\newcommand{\varNcells}{\texttt{Ncells}}
\newcommand{\varMspecies}{\texttt{Mspecies}}
\newcommand{\varNreplicas}{\texttt{Nreplicas}}
\newcommand{\varMreactions}{\texttt{Mreactions}}
\newcommand{\varNdofs}{\texttt{Ndofs}}

\newcommand{\vardsize}{\texttt{dsize}}
\newcommand{\varD}{\texttt{D}}
\newcommand{\varu}{\texttt{u0}}
\newcommand{\varN}{\texttt{N}}
\newcommand{\varG}{\texttt{G}}
\newcommand{\varK}{\texttt{K}}
\newcommand{\varI}{\texttt{I}}
\newcommand{\varS}{\texttt{S}}
\newcommand{\vartspan}{\texttt{tspan}}

\newcommand{\varreport}{\texttt{report}}
\newcommand{\varvol}{\texttt{vol}}
\newcommand{\varsd}{\texttt{sd}}

\newcommand{\varsolverargs}{\texttt{solverargs}}
\newcommand{\varmakeargs}{\texttt{makeargs}}
\newcommand{\varldata}{\texttt{ldata}}
\newcommand{\vargdata}{\texttt{gdata}}

\newcommand{\varurdmesd}{\texttt{urdme\_sd}}
\newcommand{\varurdmesdlevel}{\texttt{urdme\_sdlevel}}

\newcommand{\varU}{\texttt{U}}

\newcommand{\Ncells}{N_{\mbox{\footnotesize{cells}}}}

\newcommand{\Mreactions}{M_{\mbox{\footnotesize{reactions}}}}

\newcommand{\rand}{\mbox{rand}}

\newcommand{\fatx}{\mathbf{x}}

\title{The URDME manual \\ \large{Version 1.5}}

\author{Stefan Engblom$^{\mbox{{\tiny 1}}}$ \textit{et al.}$^{\mbox{{\tiny 2}}}$}

\date{\today}

\begin{document}
\selectlanguage{english}

\maketitle

{\footnotesize \em $^{\mbox{\tiny\rm 1}}$ Division of Scientific
  Computing, Department of Information Technology, Uppsala University,
  P.~O.~Box 337, SE-75105 Uppsala, Sweden, e-mail:
  \href{mailto:stefane@it.uu.se}{stefane@it.uu.se}.}

{\footnotesize \em $^{\mbox{\tiny\rm 2}}$ For a full list of authors,
  see the file \href{https://github.com/URDME/urdme/blob/master/AUTHORS}{AUTHORS}.}


\medskip

{\footnotesize \textbf{Manager of this release:} Stefan Engblom (to
  whom correspondence should be addressed).}

\section{Introduction}
\label{sec:intro}

Stochastic simulation methods are frequently used to study the
behavior of cellular control systems modeled as continuous-time
discrete-space Markov processes (CTMC). Compared to the most
frequently used deterministic model, the reaction rate equations, the
mesoscopic stochastic description can capture effects from intrinsic
noise on the behavior of the networks \cite{McAdams:1999fk,
  BARKAILEIBLER, Elowitz:2002fj, THATTAI, Paulsson:2000kx}.

In the discrete mesoscopic model the state of the system is the copy
number of the different chemical species and the reactions are usually
assumed to take place in a well-stirred reaction volume. The chemical
master equation is the governing equation for the probability density,
and for small to medium sized systems it can be solved by direct,
deterministic methods \cite{MacThesis, Ferm:fr, SPGRIDS1, EngSpect1,
  EngSpect2}. For larger models however, exact or approximate kinetic
Monte Carlo methods \cite{SSA, TAULEAP} are frequently used to
generate realizations of the stochastic process. Many different hybrid
and multiscale methods have also emerged that deal with the typical
stiffness of biochemical reactions networks in different ways, see
\cite{MSSA, nSSA, MYHYB, Rao:2003df, haseltine:6959} for examples.

Many processes inside the living cell can not be expected to be
explained in a well-stirred context. The natural macroscopic model is
the reaction-diffusion partial differential equation equation (PDE)
which has the same limitations as the reaction rate equations. By
discretizing space with small subvolumes it is possible to model the
reaction-diffusion process by a CTMC within the same formalism as for
the well-stirred case. A diffusion event is now modeled as a first
order reaction from a subvolume to an adjacent one and the state of
the system is the number of molecules of each species in each
subvolume. The corresponding master equation is called the
reaction-diffusion master equation (RDME) and due to the very high
dimensionality it cannot be solved by deterministic methods for
realistic problem sizes.

The RDME has been used to study biochemical systems in \cite{BISTAB,
MinD}. Here the next subvolume method (NSM) \cite{BISTAB}, an
extension of Gibson and Bruck's next reaction method (NRM) \cite{NRM},
was suggested as an efficient method for realizing sample
trajectories. An implementation on a structured Cartesian grid is
freely available in the software MesoRD~\cite{mesoRD}.

The method was extended to unstructured meshes in \cite{SPDEPEFHL} by
making connections to the finite element method (FEM). This has
several advantages, the most notable one being the ability to handle
complicated geometries in a flexible way. This is particularly
important in cell-biological models where internal structures often
must be taken into account.

This manual describes the software URDME which provides an efficient,
modular implementation capable of stochastic simulations on
unstructured meshes. URDME is easy to use for simulating and studying
a particular model in an applied context, but also for developing and
testing new approximate methods. We achieve this by relying on
third-party software for the geometry definition, meshing,
preprocessing and visualization, while using a highly efficient
computational core written in ANSI C for the actual stochastic
simulation. This keeps the implementation of the Monte Carlo part
small and easily expandable, while the user benefits from the advanced
pre- and post-processing capabilities of modern FEM software. In this
version of URDME, we maintain an older interface to Comsol
Multiphysics \cite{COMSOL}, but also to the software PDE Toolbox
\cite{PDE}, often available with standard Matlab university licenses.
        
The rest of this manual is organized as
follows. Section~\ref{sec:release} summarizes the major changes of the
1.5 release as well as the downloading and installation procedures. An
overview of the software structure is presented in
Section~\ref{sec:ov2} and the details concerning the input to the
code, the provided interface to Comsol and the way models should be
specified are found in Section~\ref{sec:details}. An URDME model is
set up and simulated in a step-by-step manner in Section~\ref{sec:ex}
and in Section~\ref{sec:integration} we explain how a new solver can
be integrated into the URDME infrastructure.

In two appendices we recapitulate the mesoscopic reaction-diffusion
model and show how the stochastic diffusion intensities are obtained
from a FEM discretization of the diffusion equation. We also list for
reference a few stochastic simulation algorithms.

We refer the interested reader to the earlier paper \cite{URDMEpaper}
for further information on the URDME software, including comparisons
to other available software and examples of some more advanced usage.

\section{The URDME~1.5 release in short}
\label{sec:release}

The major changes compared to URDME~1.4~\cite{URDMEman5} are as
follows:
\begin{enumerate}
\item The format of the URDME structure has been updated. For a full
  definition of this format, type \texttt{type urdme}.

\item Minor improvements to the solvers have been made. For example,
  the URDME structure fields \texttt{data\_time}, \texttt{ldata\_time}
  and \texttt{gdata\_time} explicitly allow for models with time
  dependent data.

\item A new function \texttt{mexjac} allows for analytical Jacobians
  with respect to the state variables to be computed. The data
  required for this function can be obtained from \texttt{rparse}.

\item The workflow \texttt{DLCM}, for modeling of populations of cells
  \cite{DLCM_solver, laplace_cellmech, algNDR} has been completely
  revised and now constitutes a specific solver for cell population
  models. Please refer to the
  \href{https://github.com/URDME/urdme/blob/master/workflows/DLCM/README.md}{README}
  for more information.
\end{enumerate}

\begin{description}
\item[Downloading/installing] There is no install procedure, simply
  download URDME 1.5 and after calling the \texttt{startup}-function
  for setting the paths you are ready to simulate.

\item[System requirements] Please refer to the file
  \href{https://github.com/URDME/urdme/blob/master/VERSION}{VERSION},
  which lists the platforms and Matlab and PDE Toolbox versions we
  have tested. There are additional minor dependencies not listed
  there which might affect certain parts of URDME.
  
\item[License] URDME is work in progress. You may use, distribute, and
  modify the code freely under the GNU GPL license version 3. We
  welcome contributions, suggestions, comments, and bug-reports. Refer
  to the file
  \href{https://github.com/URDME/urdme/blob/master/LICENCE}{LICENCE}.

\item[Quick start] Navigate to the examples folder and pick one of the
  examples. Some of the examples will benefit from PDE Toolbox or a
  live Comsol connection.

\item[Web-page] \url{http://www.urdme.org}.
\end{description}

\section{Software overview}
\label{sec:ov2}

URMDE consists of three logical layers. The top layer is made up of an
interface to an external PDE solver and pre-/post-processing engine,
currently Comsol Multiphysics or PDE Toolbox. The use of this layer is
to define diffusion- or transport rates. The bottom layer is a set of
simulation routines written in a compiled language (typically
C). Interfacing those two levels is a middle layer in the Matlab
environment, designed to facilitate data processing and visualization,
as well as custom model development. Together these layers form a
software package that enables powerful development and efficient
simulation of complex models of spatial stochastic phenomena.

The URDME structure is designed with both efficiency and flexibility
in mind. In Matlab, an URDME model is defined by a single entity,
\emph{the URDME structure}, conventionally called \umod. This
structure carries all information about the model to be
simulated. Indeed, the most bare use of \urdme\ is
\begin{verbatim}
>> umod = urdme(umod);
\end{verbatim}
Depending on the contents of \umod, this call typically compiles the
propensity source file defined in the field \texttt{umod.propensities}
using the \mex-based compilation script named
\texttt{make\_$\langle$umod.solver$\rangle$.m}. After compilation the
solver \texttt{mex$\langle$umod.solver$\rangle$} is called with
arguments formed from the fields of \umod\ and the result of the
simulation is attached to the field \texttt{umod.U}. A more explicit
call achieving the same thing is
\begin{verbatim}
>> umod = urdme(umod,'propensities',<file>,'solver',<solver>);
\end{verbatim}
Type \texttt{help urdme} for more information on the different options
available.  Type \texttt{type urdme} for a definition of the URDME
structure.

A model is conveniently built in three separate steps, one for each of
the logical layers. For example, the geometry of the model can be
defined in a Comsol \texttt{.mph} model file, along with the names and
diffusion rates of each chemical species. A Matlab model file supplies
the model with the stoichiometric matrix, the dependency graph, and
the initial state of the system. Briefly, the stoichiometric matrix
defines the effect of the chemical reactions on the state of the
system while the dependency graph indicates the reaction rates that
need to be updated after a given reaction or diffusion event has
occurred. Finally, a model file written in a compiled language
specifies the propensity functions for the chemical reactions in the
system. Using compiled rather than interpreted reaction rates ensures
maximum efficiency when simulating the model. Alternatively, for
mass-balance kinetics, the very efficient inline propensities may be
used instead.

\subsection{The modeling steps in some more detail}
\label{subsec:detailed_steps}

The steps involved in performing a URDME simulation is outlined below,
along with the routines that perform the different tasks.
\begin{enumerate}
\item Process the \texttt{.mph} model file. This is achieved by
  loading the Comsol Java object into the Matlab workspace and
  invoking the routine \texttt{comsol2urdme}. This initializes the
  \umod-structure with various fields and additionally stores the
  original Comsol Java object in the field \texttt{umod.comsol}. The
  \umod-structure contains the fields \varD, \varvol, and \varsd,
  i.e., those data structures related to the geometry of the model and
  to the unstructured mesh. See Table~\ref{tab:fields}.

  As a somewhat less advanced alternative, PDE Toolbox may be used
  instead of Comsol for this step. See \texttt{pde2urdme}.

\item The next step is to use Matlab to initialize the remaining
  essential data fields, \vartspan, \varu, \varN, and \varG. Optional
  fields include \varldata, \vargdata, \varsolverargs, and
  \varmakeargs. They should all be added as fields to \umod. Any
  modifications of the data structures added to the model by
  \texttt{comsol2urdme} in the previous step is typically performed in
  this step as well. Again, see Table~\ref{tab:fields}.

  The reaction propensities need to be specified, either as data
  matrices in a certain format (``inline propensities''), or as source
  code to be compiled. Consider using the utility functions
  \texttt{rparse} or \texttt{rparse\_inline} at this stage.

\item After \umod\ is complete, \urdme\ is called. Some additional
  arguments may be parsed directly to \urdme\ as
  property/values. \urdme\ will call the function
  \texttt{urdme\_validate} to perform error-checking on the input to
  make sure that all required fields in \texttt{umod} are present and
  have the correct properties. Finally, the chosen solver and
  propensities are compiled using \mex\ and then called by \urdme.

\item After successful simulation, the resulting trajectory is written
  to the field \texttt{umod.U}. This can then be transferred back to
  Comsol for visualization via the function \texttt{urdme2comsol} (for
  an alternative, see \texttt{urdme2pde}).
\end{enumerate} 

Table~\ref{tab:code} shows the directory structure of URDME together
with a short description of the most important
routines. Table~\ref{tab:fields} similarly lists the fields of the
URDME structure.

\begin{table}[htp]
  \centering
  \begin{tabular}{|l|l|p{0.55\linewidth}|}
    \hline
    {\bf{Directory}} & {\bf{File(s)}} & {\bf{Description}} \\
    \hline
    & &\\
    comsol & comsol2urdme.m 
    & Matlab function converting Comsol's Java object to an initial 
    \urdme-struct \umod. \\
    & urdme2comsol.m & Matlab function for conversion of the output of
    \urdme~to the solution format used by Comsol. \\
    & & \\
    pde & pde2urdme.m 
    & PDE Toolbox interface routines, as above. \\
    & urdme2pde.m & \\
    & & \\
    doc & manual.pdf & The most recent version of this manual. \\
    & & \\
    include & binheap.h & Binary heap for managing events in some solvers. \\
    & inline.h & Declaration of the inline propensity function. \\
    & propensities.h & Definition of the propensity function datatype.\\
                     & & \\
    msrc  & urdme.m & Main solver routine. \\
                     & urdme\_validate.m & Input validation. \\
    & urdme\_validate\_model & Model validation. \\
    msrc/utils & rparse.m & Propensity C-code generation. \\
                     & rparse\_inline.m & Inline propensity data generation. \\
    & & \\
    src & binheap.c & Implementation of the binary heap. \\
    & inline.c & Inline propensity function definition. \\
      &	propensities.c & Empty, used for empty propensity source. \\
    src/nsm &  & Included solvers: NSM \cite{BISTAB} is the default solver. \\
    src/aem &  & All events method \cite{AEM}. \\
    src/uds &  & URDME Deterministic solver \cite{master_moment}. \\
    src/ssa &  & For help on one solver, type e.g., \texttt{help ssa}.\\
    src/dlcm &  & DLCM solver \cite{DLCM_solver}. \\
                     & & \\
    examples & (various) & See Section~\ref{sec:ex}.\\
    & & \\
    \hline
  \end{tabular}
  \caption{Overview of some of the files that make up URDME.}
  \label{tab:code}
\end{table}

\begin{table}
  \begin{small}
\begin{tabular}{|l|p{0.3\linewidth}|p{0.45\linewidth}|}
  \hline
  {\bf{Name}} & {\bf{Type}} & {\bf{Description}}			\\
  \hline
  \vartspan &  Vector          & A sequence of points in 
  time where the state of the system is to be returned.			\\
  \varu & Matrix [\varMspecies$\times$\varNcells\ $\times$\varNreplicas]
                            & \varu$(i,j,k)$ 
  is the initial number of species $i$ in subvolume $j$ for replica $k$. 		\\
  \varD & Sparse matrix [\varNdofs $\times$ \varNdofs], 
  where the total number of degrees of freedom is \varNdofs\ = 
  \varMspecies $\times$ \varNcells
    & The \emph{transpose} of the diffusion matrix $M^{-1}K$ obtained 
    from the FEM discretization of the macroscopic diffusion equation, 
    cf.~\eqref{eq:FEM}. Each column in \varD\ (i.e.~each row in 
    $M^{-1}K$) corresponds to a subvolume, and the non-zero coefficient 
    $D(j,i)$ gives the diffusion rate constant from subvolume $i$ to 
    subvolume $j$. 							\\
  \varN & Sparse matrix [\varMspecies\ $\times$ \varMreactions]       
    & The stoichiometric matrix. Each column corresponds to a reaction, 
    and execution of reaction $j$ amounts to adding the 
    $j$th column to the state vector.					\\
  \varG & Sparse matrix [\varMreactions\ $\times$ 
    (\varMspecies+\varMreactions)]       
    & Dependency graph. The first \varMspecies\ columns correspond to 
    diffusion events and the following \varMreactions\ columns to 
    reactions. A non-zeros entry in element $i$ of column $j$ indicates 
    that propensity $i$ needs to be updated if the event $j$ 
    occurs. See Section~\ref{sec:ex} for examples.			\\
  \varvol & Vector [\varNcells]           
    & The volume of the macroelements, i.e. the diagonal elements of 
    the lumped mass-matrix $M$ (cf.~Appendix~\ref{subsec:mesodiff}). \\
  \varsd &  Vector [\varNcells]       
    & The subdomain numbers of all subvolumes. See Section~\ref{sec:ex} 
    for more details.							\\
  \hline

  \texttt{solver} & String & Name of solver. \\
  \texttt{propensities} & String & Propensity source file. \\
  \varreport & Scalar & Report level, typically 0, 1, or 2. \\
  \texttt{solve} & Boolean & Solve on/off. \\
  \texttt{compile} & Boolean & Compilation on/off. \\
  \texttt{parse} & Boolean & Parsing on/off. \\
  \texttt{seed} & Vector [\varNreplicas] & Random seed value(s). \\
  \hline

  \texttt{modelname} & String & Name of model. \\
  \texttt{inline\_propensities} & Inline propensity structure &
   Definition of inline propensities. \\
  \varldata &  Matrix [\vardsize$\times$\varNcells]
  & Local data vector. A pointer to column $j$ is passed as an 
  additional argument to the propensities in subvolume $j$.		\\
  \vargdata &  Vector [$\langle$anything$\rangle$]
                            & Global data vector \\
  \texttt{data\_time} & Increasing vector of time points. & Nodes for time-dependent data. \\
  \texttt{ldata\_time} & 3D Array
                         [$\langle$anything$\rangle$ $\times$
                         \varNcells $\times$ \texttt{length(data\_time)}
                            & Local time-dependent data vector. \\
  \texttt{gdata\_time} & Matrix [$\langle$anything$\rangle$ $\times$ \texttt{length(data\_time)}] & Global time-dependent data vector. \\
  \texttt{solverargs} & Cell-vector of property/value pairs. &
  Solver arguments, must be parsed in this form by the solver
                                                               \mex-interface. \\
  \varmakeargs & Cell-vector of property/value pairs. &
  Make arguments, must be parsed in this form by the solver make-file. \\
  \hline

  \varU & Matrix [\varNcells $\times$ \texttt{length}(\vartspan)
          $\times$ \varNreplicas] & Latest stored solution. \\
  \texttt{comsol}, \texttt{pde} & Object fields & 
  Comsol Java object and PDE Toolbox data. \\
  \texttt{private} & $\langle$anything$\rangle$ & Arbitrary additional data. \\
  \hline
\end{tabular}
\caption{The fields of the URDME structure. \textit{Top:} required
  before call to \urdme, \textit{2nd:} passed as arguments to \urdme\
  with default values, \textit{3rd:} optional fields with empty
  defaults, \textit{bottom:} optional fields. Type \texttt{type urdme}
  to see this information.}
\label{tab:fields}
  \end{small}
\end{table}

\section{Details and specifications}
\label{sec:details}

In this section we give a detailed description of the input to the
URDME solvers.

\subsection{The \mex-interface}

The URDME solver sequence is readily summarized as:
\begin{verbatim}
>> mexmake_<umod.solver>(umod.propensities,umod.makeargs{:});
>> umod.U = <umod.mexexec>(umod.mexhash, ...
                           umod.tspan,umod.u0, ...
                           umod.D,umod.N,umod.G, ...
                           umod.vol,umod.ldata,umod.gdata, ...
                           umod.data_time, ...
                           umod.ldata_time,umod.gdata_time, ...
                           umod.sd, ...
                           umod.report,umod.seed, ...
                           umod.inline_propensities.K, ...
                           umod.inline_propensities.I, ...
                           umod.inline_propensities.S, ...
                           umod.solverargs);
\end{verbatim}
For the meaning of the different fields in \umod, see
Table~\ref{tab:fields}. Samples of actual source code handling the
final \mex-interface call are found in the source folders. Contributed
solvers should adhere to exactly the above \mex-interface.

\subsection{Specifying propensities for chemical reactions}

We have provided two separate methods to specify the reaction
propensities. Simple polynomial rate laws (mass-action) can be
provided as inline propensities and can be specified in the Matlab
layer. For general propensities and full flexibility, the rate laws
can be specified in a model file written in a compiled language (C
typically).

Note that one can easily use \emph{both} inline and compiled
propensities simultaneously. This might be convenient when only a few
propensities are complicated and has to be compiled.

\subsubsection{Inline propensities}

An ``inline propensity'' is a compact data format for specifying basic
chemical reactions with polynomial rate laws. An inline propensity $P_r$
can be defined as
\begin{align*}
  P_r(x) &= \left\{
 \begin{array}{rl}
    \frac{k_1x_ix_j}{\Omega}+k_2x_k+k_3\Omega & \text{if } i \ne j,	\\
    \frac{k_1x_i(x_i-1)}{2\Omega}+k_2x_k+k_3\Omega & \text{if } i = j.
  \end{array} \right.
\end{align*}
Here $x$ is the column in $\fatx$ which contains the state of the
subvolume considered and $\Omega$ is the corresponding volume. The
coefficients and indices are specified in matrices \varK\ and
\varI\ where \varK$(:,r) = [k_1 \, k_2 \, k_3]^T$ and \varI$(:,r) = [i
  \, j \, k]^T$ are the constants corresponding to the $r$th inline
propensity. The matrix \varS\ is a (possibly empty) sparse matrix such
that \varS$(:,r)$ lists all subdomains in which the $r$th inline
propensity is turned off. \emph{Note that no inline propensities are
  active in subdomain zero!} A complete example of the use of inline
propensities can be found in the 'annihilation' example folder.

\medskip \noindent {\bf \textcolor{red}{!}} \emph{The format
  specification for inline propensities might feel a bit complicated
  at first!} The utility function \texttt{rparse\_inline} can aid in
constructing these from simple expressions.

\subsubsection{Compiled propensities}

The other way to specify propensity functions is to supply them to
\urdme\ as a propensity file written in C. The precise form of the
propensity functions is defined by the data type
\texttt{PropensityFun}, defined in the header `propensities.h' (found
in the `include' directory) as

\begin{verbatim}
typedef double (*PropensityFun)(const URDMEstate_t *x,double t,double vol,
                                const double *ldata,const double *gdata,
                                int sd);
\end{verbatim}
The default type for \texttt{URDMEstate\_t} is just a plain
\texttt{int}, but the UDS-solver requires a \texttt{double}.

The arguments \varvol, \varldata, \vargdata, and \varsd\ are described
in Table~\ref{tab:fields}. Additionally, the input vector \texttt{x}
of length \varMspecies\ is the copy number in a given subvolume, and
\texttt{t} is the absolute time. Note that, of the current URDME
solvers, only UDS makes an active use of the time.
 
Below is a commented example of a model file defining a simple
chemical system composed of a single species undergoing a
dimerization reaction.

\begin{verbatim}
/* Propensity definition of a simple dimerization reaction. */
#include "propensities.h"
#include "report.h"

const int NR = 1; /* number of reactions */

const double k = 1.0e-3; /* rate constant */

/* forward declaration */
double rFun1(const int *x,double t,double vol,
             const double *ldata,const double *gdata,int sd);

/* static propensity vector */
static PropensityFun ptr[] = {rFun1};

double rFun1(const int *x,double t,double vol, 
             const double *ldata,const double *gdata,int sd)
/*  X + X --> 0. */
{
  return k*x[0]*(x[0]-1)/vol;
}

PropensityFun *ALLOC_propensities(size_t Mreactions)
{
  if (Mreactions > 1) PERROR("Wrong number of reactions.");
  return ptr;
}

void FREE_propensities(PropensityFun *ptr)
{ /* do nothing since a static array was used */ }
\end{verbatim}
\noindent
A propensity file \emph{must} implement the following routines:
\begin{itemize}
  \item \verb#PropensityFun *ALLOC_propensities(size_t Mspecies)#
  \item \verb#void FREE_propensities(PropensityFun *ptr)#
\end{itemize}
The first function should allocate and initialize an array of function
pointers to the propensity functions and return a pointer to this
array. This is the function that the solvers will call to access the
rate functions. The second function should deallocate the pointer
\texttt{ptr}, whenever this is required. In the above example, a
static array was used and deallocation is unnecessary. For further
examples, see Section~\ref{sec:ex}.

\medskip \noindent {\bf \textcolor{red}{!}} \emph{The propensity
  function specification can be a bit cumbersome at first!} Consider
starting your work with the function \texttt{rparse}. For example, the
above example is readily generated directly as
\begin{verbatim}
>> rparse({'X+X > k*X*(X-1)/vol > @'},{'X'},{'k' 1e-3});
\end{verbatim}
Note that the fields \texttt{umod.N} and \texttt{umod.G} can be
generated as well.

\section{Worked examples}
\label{sec:ex}

In this section we describe the general workflow involved in setting
up and simulating a model in URDME using the Comsol and Matlab
interfaces. The major steps are (compare
Section~\ref{subsec:detailed_steps}):
\begin{enumerate}
\item \emph{Specify the model}. This involves defining the geometry,
  mesh, initial conditions and chemical reactions of the model. In
  URDME, this is conveniently accomplished using three model files: a
  Comsol model file `model.mph', a Matlab model file `model.m' and a
  reaction propensity file `model.c', where we use {\it model} as a
  placeholder for the non-extension part of the file-name.

  \begin{enumerate}
    \item The Comsol model defines the geometry of the problem and
      Comsol Multiphysics is used to create a mesh representing the
      spatial discretization of the diffusion equation with Neumann
      boundary conditions and the inter-voxel diffusion jump
      coefficients. The Comsol model is exported to the Matlab layer,
      typically using the \texttt{mphload} command. The function
      \texttt{comsol2urdme} then extracts the information into a
      startling \umod-structure. A simpler alternative to Comsol is
      PDE Toolbox. Below we show examples of both methods.
    \item The Matlab model file specifies the chemical reaction
      networks of the problem.
    \item The propensity functions for the chemical reactions are
      either specified in a C-source code file and/or in the Matlab
      layer as inline propensities. These are typically created using
      the utility functions \texttt{rparse} and/or
      \texttt{rparse\_inline}.
  \end{enumerate}
	
\item \emph{Run the simulation.} The simulation takes place in the
  Matlab workspace via a call to \urdme.
  	
\item \emph{Post-processing.} After a normal termination of the
  solver, a trajectory of the stochastic process will be attached to
  the field \texttt{umod.U}. The function \texttt{urdme2comsol}
  attaches this trajectory to the Comsol Java object in
  \texttt{umod.comsol}. At this point, you can use all post-processing
  options available in the Comsol interface to visualize the
  results. You may also save the solution in \texttt{umod.comsol} to
  file via the command \texttt{mphsave} and later post-process it in
  the Comsol GUI. Again and as an alternative, you may rely on PDE
  Toolbox rather than on Comsol.
\end{enumerate}

\subsection{Morphogenesis: the Schnakenberg model}

As an immediate example of the simulation capabilities and general
workflow of URDME, we reproduce one of the simulations found in
\cite{Yakup_exjobb}, namely a stochastic version of the Schnakenberg
model. The model consists of two \emph{morphogens} $U$ and $V$
diffusing in some geometry and reacting according to the transitions
\begin{align}
  \begin{array}{rcl}
    \emptyset  & \xrightarrow{ k_1 \Omega } &  U	\\
    U  & \xrightarrow{ k_2 U } &  \emptyset	\\
    \emptyset  & \xrightarrow{ k_3 \Omega } &  V	\\
    U+U+V  & \xrightarrow{ k_4/\Omega^2 \, U(U-1)V } &  U+U+U	\\
  \end{array}
\end{align}

We set up and solve the model step by step as follows; see also the
file \texttt{morphogenesis2D\_run.m} in the folder
`examples/morphogenesis'.

\paragraph*{Defining the geometry and diffusion rates in PDE Toolbox}

\begin{figure}[htp]
  \centering
  \includegraphics[width = 0.75\textwidth]{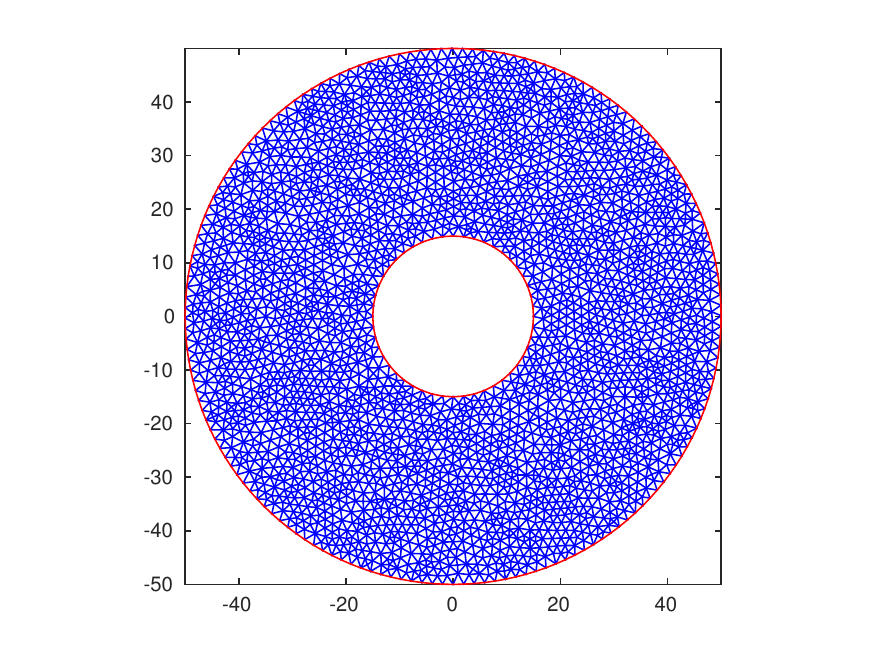}
  \label{fig:Schnakenberg_mesh}
  \caption{Mesh of the Schnakenberg model.}
\end{figure}

We start by building the geometry using the format supported by PDE
Toolbox, in this case a 2D torus:
\begin{verbatim}
C1 = [1 0 0 50]';
C2 = [1 0 0 15]';
gd = [C1 C2];
sf = 'C1-C2';
ns = char('C1','C2')';
G = decsg(gd,sf,ns);
\end{verbatim}

We next build the mesh as follows:
\begin{verbatim}
[P,E,T] = initmesh(G,'hmax',2.5);
pdemesh(P,E,T), axis tight, axis equal
\end{verbatim}
The result of the plot-command is displayed in
Figure~\ref{fig:Schnakenberg_mesh}.

The diffusion parameter is different for the two species. We use PDE
Toolbox to assemble the discretized diffusion operator:
\begin{verbatim}
D_U = 1;
D_V = 40;
umod = pde2urdme(P,T,{D_U D_V});
umod.sd = ceil(umod.sd); % (not used)
\end{verbatim}
PDE Toolbox produces an interpolated value of \varsd\ and a `decision'
is required for how to handle this value. Above we simply round the
result upwards; a model making more advanced use of the subdomain
numbering needs to handle this issue with more care.

\paragraph*{Specifying the chemical reactions}

We readily construct the propensity file using the utility-function
\texttt{rparse}:
\begin{verbatim}
r1 = '@ > k1*vol > U';
r2 = 'U > k2*U > @';
r3 = '@ > k3*vol > V';
r4 = 'U+U+V > (k4/(vol*vol))*(U*(U-1)*V) > U+U+U';
umod = rparse(umod,{r1 r2 r3 r4},{'U' 'V'}, ...
         {'k1' 0.1 'k2' 1 'k3' 0.9 'k4' 1}, ...
         'schnakenberg.c');
\end{verbatim}
The result of the above command is written on the indicated file. Take
a moment to study the resulting source code!

We also need to define a few more fields of the \umod-struct:
\begin{verbatim}
umod.u0 = zeros(size(umod.N,1),numel(umod.vol));
umod.tspan = 0:10:100;
umod.vol = 50/mean(umod.vol)*umod.vol;
\end{verbatim}
The last command is special to the current model: it is of some
interest to investigate how the Schnakenberg model responds to
different scaling of the system volume. Here this is achieved by
directly rescaling the field \varvol\ of the \umod-struct.

\paragraph*{Simulating the model}

At this stage we simply invoke \urdme:
\begin{verbatim}
umod = urdme(umod,'report',3);
\end{verbatim}
{\bf \textcolor{red}{!}} Using \varreport\ = 3 implies that you have
the possibility to discontinue the simulation at each entry in
\vartspan. A useful feature when setting up new models.

\paragraph*{Postprocessing}

To use PDE Toolbox in the visualization process we rely on the
function \texttt{urdme2pde}:
\begin{verbatim}
umod = urdme2pde(umod);
figure, clf,
pdeplot(P,E,T,XYdata=umod.pde.U(1,:,end)');
title('Schnakenberg: Concentration U');
view(0,90), axis tight, axis square, colormap('parula')
\end{verbatim}
The result of these plot-commands can be inspected by running the
example. In the file \texttt{morphogenesis2D\_run.m} a second model,
\emph{the Brusselator}, is also solved in a similar fashion.

\subsection{Min oscillations in \emph{E.~Coli}}

To illustrate the solution steps in a more advanced 3D model, we will
reproduce simulations of the Min-system from \cite{MinD}. The geometry
will be a model of an \emph{E.~coli} bacterium. It is rod-shaped with
length $3.5\mu m$ and diameter $0.5\mu m$. The reactions and
parameters of the model can be found in Table~\ref{tab:MinD}. The
model is built in the file \texttt{mincde\_run.m} found in the folder
`examples/mincde'.

\begin{table}[htp]
\centering
\begin{tabular}{ll}
  $\mbox{MinDcytATP} \xrightarrow{k_d} \mbox{MinDmem}$ & 
  $\mbox{MinDcytATP}+\mbox{MinDmem} \xrightarrow{k_{dD}} \mbox{2MinDmem}$ \\
  $\mbox{MinE+MinDmem} \xrightarrow{k_{de}} \mbox{MinDE}$ & 
  $\mbox{MinDE} \xrightarrow{k_e} \mbox{MinDcytADP}+\mbox{MinE}$ \\
  $\mbox{MinDcytADP} \xrightarrow{k_p} \mbox{MinDcytATP}$
 \end{tabular} 
  \caption{The chemical reactions of the Min system. The constants
    take the values $k_d = 0.0125\mu m^{-1}s^{-1}$, $k_{dD} = 9 \times
    10^6 M^{-1}s^{-1}$, $k_{de} = 5.56\times 10^7 M^{-1}s^{-1}$,
    $k_e=0.7s^{-1}$, and $k_p = 0.5s^{-1}$.}
  \label{tab:MinD}
\end{table}

\subsubsection{Setting up the model for simulation}

\paragraph*{Defining the geometry and diffusion rates in Comsol Multiphysics}

\begin{enumerate}
\item Open Comsol and use the Model Wizard to create the model
  template. Select `3D' as space dimension and add the physics module
  `Chemical Species Transport / Transport of Diluted Species' in the
  next step. In the `Dependent variables' window chose the `Number of
  species' to be \texttt{5} and in the `Concentrations' list enter the
  names \texttt{MinDcytATP, MinDmem, MinE, MinDE} and
  \texttt{MinDcytADP}. You may also enter these variable names at a
  later stage, see below. Select the `Time Dependent' study type in
  the next step of the wizard and click on the flag symbol to create
  the template.
  
  {\bf \textcolor{red}{!}} \emph{Note that the `Chemical engineering
    module' is not required in general for URDME, but is used in this
    example for convenience.}

\item Next we create the geometry. We will build the rod shaped domain
  from two spheres and one cylinder. Right click on `Geometry 1',
  select the `Cylinder' option and in the radius and height field
  enter \texttt{0.5e-6} and \texttt{3.5e-6}. Click on the `Build
  Selected' Button and you should now see a cylinder in your
  workspace. Now, select the `Sphere' node from the `Geometry'
  context-menu and and enter \texttt{0.5e-6} in the radius
  field. Create another identical sphere but enter \texttt{3.5e-6} as
  the $x$-coordinate. Click on `Build All' and observe the created
  domain in the graphics-window.
  
  Right click on `Geometry' again and select `Boolean Operations $>$
  Union'. Select all three domains and add them to the `Input objects'
  selection. Uncheck the `Keep interior boundaries' box and complete
  the geometry creation by pushing the `Build All' button. The final
  geometry has 1 domain, 12 boundaries, 20 edges, and 10 vertices.

\item Having constructed the geometry, the next step is to specify the
  parameters in the model. If you haven't specified the variable names
  yet, do this now under `Transport of Diluted Species $>$ Dependent
  Variables' (Number of species: 5, Concentrations:
  \texttt{MinDcytATP}, \texttt{MinDmem}, \texttt{MinE},
  \texttt{MinDE}, \texttt{MinDcytADP}).

  In the physics settings `Transport of Diluted Species $>$ Transport
  Mechanisms', deactivate the flag on `Convection'. Also, with the
  `Transport of Diluted Species' tab active, press the advanced
  property icon (an all-seeing eye just under the `Model Builder'
  title), and select `Discretization'. Use linear elements and uncheck
  the `Compute boundary fluxes'.

  Next we need to specify the diffusion constants of the species in
  the `Diffusion' node of the physics menu `Transport
  Properties'. Enter the diffusion coefficients \texttt{2.5e-12} for
  \texttt{MinDcytATP}, \texttt{MinE}, and \texttt{MinDcytADP}. For
  \texttt{MinDE} and \texttt{MinDmem} the diffusion constant should be
  \texttt{1e-14}. The units of all constants are $m^2/s$.
  
  \smallskip 
   
  {\bf \textcolor{red}{!}} \emph{\texttt{MinDE} and \texttt{MinDmem}
    are species bound to the membrane, hence their lower diffusion
    rates. We have not specified this explicitly at this stage, but
    will do so later in the Matlab layer.}

\item In order to be able to distinguish between the interior of the
  bacterium and the membrane, we must also create two domains which
  URDME can parse. One interior domain that represents the cytoplasm
  and one boundary domain that represents the membrane. This is done
  by (1) defining the global \varurdmesdlevel\ variable, and (2)
  defining the variable \varurdmesd\ on the geometry as an expression
  with different values in the different subdomains. The latter
  variable is then used by URDME to distinguish the nodes on the
  boundary from those in the interior. Somewhat technically, the
  former variable, \varurdmesdlevel, is required to correctly evaluate
  the value of \varurdmesd\ on the different manifolds of the model
  (points, edges, boundaries, domain).

  First click right on `Global Definitions' and create the `Variable'
  \varurdmesdlevel. Assign this variable the value 2 to indicate that
  the \emph{lowest} dimension where \varurdmesd\ is defined is on the
  boundaries of the geometry (where the dimensionality is 2).

  Second, click right on the menu `Definitions', and create two
  `Variables'. Label the first one \texttt{urdme\_sd\_dom} (name is
  not critical) and select the `Geometry entity level' to be `Domain'
  and chose the `Selection' to be `All domains'. Now, enter a new
  variable in the window below by specifying the name to
  \varurdmesd\ and expression to \texttt{1}. Similarly, label the
  second variable \texttt{urdme\_sd\_bnd} (again, name is not
  critical), specify the geometric entity level to `Boundary' and set
  the `Selection' to `All boundaries'. Enter the variable name
  \varurdmesd\ into the `Variables' window and set the expression to
  \texttt{2}.

  All in all we have now defined the variable \varurdmesdlevel\ to be
  2 globally such that URDME will correctly evaluate the variable
  \varurdmesd\ first at the entire domain of the model (where
  \varurdmesd\ = 1) and second at the boundaries of the model (where
  \varurdmesd\ = 2). The URDME convention here is that lower
  dimensional manifolds take precedence over higher dimensional ones.

\item In the `Mesh' node, set `User controlled mesh' as sequence type
  and in the appeared `Size' node select the `Custom' option.  Set the
  maximum element size to \texttt{1e-7} and press `Build All'.  Now
  click on the `Study' node and press the `Compute' button.
  
\item Some Comsol callback functions require a placeholder
  solution. To create a simple such solution, under `Study 1: Time
  Dependent', select `Times: range(0,1,1)' and press the `Compute'
  button.
\end{enumerate}

Now you need to transfer the created model into Matlab. Make sure that
you are connected to the Server, if not, connect via `File $>$ Client
Server $>$ Connect to Server'.  When having a working connection the
export can be performed by selecting `File $>$ Client Server $>$
Export Model to Server'.
  
Another option is to save the model to file (here: \texttt{coli.mph}),
and open it later in a running Matlab session with `LiveLink' via the
command \texttt{mphload}:
\begin{verbatim}
model = mphload('coli.mph'); % load Comsol model
umod = comsol2urdme(model);  % create URDME struct
\end{verbatim}

\paragraph*{Specifying the chemical reactions}

The chemical reactions are specified in a separate file written in C,
conveniently generated on the fly by the utility function
\texttt{rparse}. Open the file `examples/mincde/mincde\_run.m'. We will
walk through the contents of this file and explain what the different
parts do. Additional information can also be found in the comments in
the file.

First out is the membrane bound reaction,
\begin{verbatim}
r1 = 'MinDcytATP > sd == 2 ? kd*MinDcytATP/ldata[0] : 0.0 > MinDmem';
\end{verbatim}
Note how \varsd\ is used to check if the voxel belongs to the membrane
or not. We have to make sure, however, that we keep track of what
value we assigned to the different subdomains in the Comsol model file
(the value of the expression \varurdmesd).

Note also how the first reaction in the model contains a scaling with
the local length scale of the subvolume. For a uniform Cartesian mesh
this would simply have been the (constant) side lengths of the cubes
in the mesh. For the unstructured mesh however, this value will be
different in every subvolume. It is readily obtained from Comsol, and
is passed to the propensity function via the data vector
\varldata\ which will be initialized with the correct values below.

We continue with the two bimolecular reactions,
\begin{verbatim}
r2 = ['MinDcytATP + MinDmem > kdD*MinDcytATP*MinDmem/(1000.0*NA*vol)' ...
      '> MinDmem+MinDmem'];
r3 = 'MinE + MinDmem > kde*MinE*MinDmem/(1000.0*NA*vol) > MinDE';
\end{verbatim}
Note the unit conversions given explicitly in the bimolecular
propensity function. The rate constants for the bimolecular reactions
in this model are given in the unit $M^{-1}s^{-1}$ and need to be
converted to mesoscopic rates. That is why we divide with Avogadros
number times the volume of the subvolume. Also, the way we have set up
the geometry model file, the volume is given in the unit $m^3$, and
needs to be converted to $L^3$. URDME cannot keep track of matching
the units between the different model files automatically: this is the
responsibility of the end-user.

The chemical network is concluded with two degradation reactions:
\begin{verbatim}
r4 = 'MinDE > ke*MinDE > MinDcytADP + MinE';
r5 = 'MinDcytADP > kp*MinDcytADP > MinDcytATP';
\end{verbatim}

To create the propensity file we define the species and the rate
constants and then invoke \texttt{rparse}:
\begin{verbatim}
species = {'MinDcytATP' 'MinDmem' 'MinE' 'MinDE' 'MinDcytADP'};
rates = {'NA' 6.022e23 'kd' 1.25e-8 'kdD' 9.0e6 'kde' 5.56e7 ...
         'ke' 0.7 'kp' 0.5};
umod = rparse(umod,{r1 r2 r3 r4 r5},species,rates,'fange.c');
\end{verbatim}
The propensity C-file is now found in the file \texttt{fange.c}. Take
a moment to study this file!

Before we can run the simulation, we need to modify the diffusion
rates that we obtain from the initial Comsol model so that the
membrane-bound species only diffuse on the membrane. We have already
prepared for this by labeling the subvolumes next to the boundary
using the expression \varurdmesd\ in the Comsol model.

\begin{enumerate}
\item \emph{The initial condition.} There is complete freedom in
  specifying the initial condition. In the present case we simply
  distribute 4002 \texttt{MinDcytATP} and 1040 \texttt{MinE} molecules
  in some random way in the entire bacterium.
\begin{verbatim}
% the total number of molecules of the species
nMinD = 4002;
nMinE = 1040;

% assign randomly
u0 = zeros(Mspecies,Ncells);
ind = floor(Ncells*rand(1,nMinE))+1;
u0(3,:) = full(sparse(1,ind,1,1,Ncells));
ind = floor(Ncells*rand(1,nMinD))+1;
u0(5,:) = full(sparse(1,ind,1,1,Ncells));
umod.u0 = u0;
\end{verbatim}
  Note that the code above does not produce a uniformly random initial
  distribution since the volume of each voxel is not taken into
  account.

\item \emph{Specifying the times to output the state of the system.}
  URDME will look for a vector \texttt{tspan} to determine when to
  output the state of the trajectory (the number of events generated
  in a typical realization often exceeds $10^9$ so we can't output
  after each event). Here, we want to sample the system on the time
  interval $[0,200]$ seconds, with output each second. This is
  achieved by
\begin{verbatim}
umod.tspan = 0:200. 
\end{verbatim}
 
\item \emph{Membrane diffusion}. In order to make \texttt{MinDmem} and
  \texttt{MinDE} diffuse only on the membrane, we will zero out all
  elements in the diffusion matrix that are in the cytosol. To obtain
  indices of those subvolumes we use the information in the subdomain
  vector \varsd.
\begin{verbatim}
cyt = find(umod.sd == 1);
pm  = find(umod.sd == 2);
\end{verbatim} 
 
  Remember that we gave \varurdmesd\ the value 2 on the membrane and 1
  in the interior. The diffusion matrix \texttt{D} will contain the
  rate constants for the diffusive events on the unstructured
  mesh. \texttt{D} is generated by Comsol and is available in the
  field \texttt{umod.D}. To (efficiently) zero out the correct entries
  in \texttt{D}, we first decompose the sparse matrix, find the
  entries using \texttt{pm} and \texttt{cyt} above, and then
  reassemble the matrix again (compensating for the removed entries by
  adjusting the diagonal of the matrix).  All in all, the code to do
  this is as follows:

\begin{verbatim}
% For MinDmem (2) and MinDE (4), flag all dofs in the cytosol for
% removal.
ixremove = [];
for s = [2 4]
  ixremove = [ixremove; Mspecies*(cyt-1)+s];
end

% Decompose the sparse matrix. 
D = umod.D';
[i,j,s] = find(D);

% Set all elements in the diffusion matrix corresponding 
% to the cytosol to zero.
ixremove = [find(ismember(i,ixremove)); find(ismember(j,ixremove))];
i(ixremove) = [];
j(ixremove) = [];
s(ixremove) = [];

% Reassemble the sparse matrix and adjust the diagonal entries. 
ixkeep = find(s > 0);
D = sparse(i(ixkeep),j(ixkeep),s(ixkeep),Ndofs,Ndofs);
d = full(sum(D,2));
D = D+sparse(1:Ndofs,1:Ndofs,-d);
umod.D = D';
\end{verbatim}

 {\color{red}{\bf !}} It is of fundamental importance that the columns
 of \texttt{D} sum to zero, and that all off-diagonal entries are
 positive. For an introduction to how \texttt{D} is constructed, see
 Appendix \ref{app:coeffs}. For a detailed account, consult
 \cite{SPDEPEFHL}.
 
 {\color{red}{\bf !}} The way we have modeled membrane diffusion is
 simply by saying that the subvolumes closest to the membrane
 constitute the membrane layer. As the mesh becomes finer near the
 boundary, the thickness of this layer will decrease, eventually
 approaching a 2D model of the membrane. One can also think of other
 ways of modeling the membrane diffusion.
 
\item \emph{The local data vector.} Finally, we need to set
  \texttt{umod.ldata} to contain the values of the length parameter
  for the subvolumes (it is needed in the first reaction). To do this,
  we use the built-in Comsol function \texttt{mphinterp} which can be
  used to evaluate an expression in any point in the domain. Here, we
  simply get the subvolume sizes by using the pre-defined expression
  \emph{h}, evaluated in the vertices of the mesh.

Relying on Comsol Multiphysics:
\begin{verbatim}
xmi = mphxmeshinfo(umod.comsol);
umod.ldata = mphinterp(umod.comsol,'h','coord', ...
                       xmi.dofs.coords(:,1:Mspecies:end),'solnum',1);
umod.ldata = umod.ldata(xmi.dofs.nodes(1:Mspecies:end)+1);
\end{verbatim}

{\color{red}{\bf !}} For more details concerning the internal ordering
of the dofs, consult the Comsol user's manual. The interface routines
\texttt{comsol2urdme} and \texttt{urdme2comsol} also contain useful
information on this matter.
\end{enumerate}

\subsubsection{Running the simulation}

With the model set up correctly, we are now ready to simulate:
\begin{verbatim}
umod = urdme(umod,'report',2);
\end{verbatim}

URDME will now compile the solver with linking to the propensities
specified in \texttt{fange.c}, and then execute the solver. The result
of the simulation is stored in \texttt{umod.U}. To prepare for
post-processing we can transfer this result back to the Comsol object,
\begin{verbatim}
umod = urdme2comsol(umod);
\end{verbatim}

\subsubsection{Post-processing}

If the simulation in the previous step completed without errors, the
model structure will now contain a realization of the stochastic
process. To visualize the trajectory, we can use any of the
visualization options available in Comsol or we can create routines of
our own. To look at the \texttt{MinDmem} distribution on the membrane
at the final time we can use Comsol's post-processing functionality.

This command creates a plot-container for the visualization to follow:
\begin{verbatim}
umod.comsol.result.create('res1','PlotGroup3D');
\end{verbatim} 

To visualize the result at a specific time, e.g., after 100s:
\begin{verbatim}
umod.comsol.result('res1').set('t','100');
\end{verbatim}

\medskip \noindent {\color{red}{\bf {!}}} \emph{You can specify any
  time in the interval you simulated, but if you specify a time that
  lies between two points in \texttt{tspan} Comsol will do
  interpolation to approximate the result at that point.}

To visualize the result of the simulation on the surface we can use:
\begin{verbatim}
umod.comsol.result('res1').feature.create('surf1','Surface');
umod.comsol.result('res1').feature('surf1').set('expr','MinDmem');
mphplot(umod.comsol,'res1');
\end{verbatim}
Where we can replace the string \texttt{'MinDmem'} with the name of
any other species.

To visualize the solution inside the domain, we need to first create a
new plot container.
\begin{verbatim}
umod.comsol.result.create('res2','PlotGroup3D');
umod.comsol.result('res2').set('t','100');
\end{verbatim} 

Now we can visualize the solution on a `slice' of the $zx$-axis of the model.
\begin{verbatim}
umod.comsol.result('res2').feature.create('slc2','Slice');
umod.comsol.result('res2').feature('slc2').set('expr','MinDcytATP');
umod.comsol.result('res2').feature('slc2').set('quickplane','zx');
umod.comsol.result('res2').feature('slc2').set('quickynumber','1');
mphplot(model,'res2');
\end{verbatim} 

\begin{figure}[htp]
  \centering
  \includegraphics[width = 0.45\textwidth]{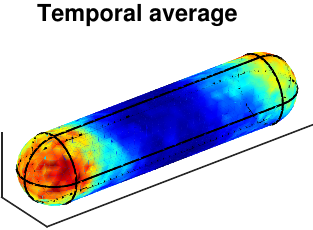}
  \includegraphics[width = 0.45\textwidth]{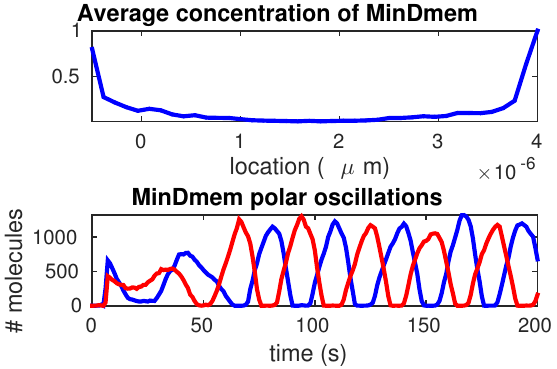}
  \label{fig:min_plots}
  \caption{Visualizations from the Min system.}
\end{figure}

There are many more options that can be passed to \texttt{mphplot} to
control the plot produced. For a detailed account, see the Comsol
documentation:
\begin{verbatim}
>> help mphplot
\end{verbatim}

If you prefer to work within the Comsol GUI for visualization, you can
import back the Comsol model with the attached stochastic trajectory
into Comsol. This can be done by typing:
\begin{verbatim}
>> mphsave(umod.comsol,'<output_filename>.mph')
\end{verbatim}

Optionally, from the Comsol GUI, import the new structure
(\texttt{umod.comsol}): 'File $>$ Client Server $>$ Import Model from
Server'. You can now visualize the trajectory using the options
provided in the `Results' node.

\section{Integrating solvers with URDME}
\label{sec:integration}

URDME is easily enriched with additional third party solvers. URDME
plugins have three main components: the \mex-based makefile, the
\mex-interface source code, and (conventionally) a placeholder Matlab
.m-file with defining help-text. Each part is described in
Table~\ref{tab:integration} where the files that make up the NSM
solver are explained. We recommend that developers follow this format
when integrating their own solvers.

\begin{table}[htb!]
  \centering
  \begin{tabular}{|l|p{0.55\linewidth}|}
    \hline
    {\bf{File}} & {\bf{Description}}\\
    \hline
     & \\
    mexmake\_nsm.m & A \mex-based makefile for building the 
    solver. The name of this file is important: the automatic 
    compilation process looks for a makefile that is suffixed with 
    the name of the solver. This makefile compiles 
    the solver with the model's propensity functions into the 
    executable \texttt{mexnsm} which is called by the Matlab-level 
    interface. \\
    & \\
    mexnsm.c & Solver entry point, a \texttt{mexFunction}. \\
    & \\
    nsm.h, nsm.c & Actual low-level source code implementing the NSM solver. \\
    & \\
    nsm.m & Empty Matlab .m-file containing help-text. \\
    & \\
    \hline
  \end{tabular}
  \caption{Overview of the files that make up the NSM solver.}
  \label{tab:integration}
\end{table}

\section*{Acknowledgment}

The development of URDME was partially funded by the Linnaeus center
of excellence UPMARC, Uppsala Programming for Multicore Architectures
Research Center, and by eSSENCE, a Swedish strategic research
programme in e-Science.

\bibliographystyle{abbrv}
\bibliography{manual}
\clearpage

\appendix
 \section{Stochastic chemical kinetics}
\label{app:coeffs}
 
In this section we briefly describe how reaction and diffusion events
are modeled and how we obtain the diffusion rate constants when the
domain is discretized using an unstructured mesh. For a more detailed
introduction to the subject along with many additional references,
consult, e.g., \cite{stefan_phd}.

The computational core of URDME is based on the next subvolume method
(NSM) \cite{BISTAB}. Details
concerning the actual simulation algorithms can be found in
Appendix~\ref{app:algorithms}.

\subsection{Mesoscopic chemical kinetics}

In a well-stirred chemical environment reactions are understood as
transitions between the states of the integer-valued state space
counting the number of molecules of each of $D$ different species. The
intensity of a transition is described by a \emph{reaction propensity}
defining the transition probability per unit of time for moving from
the state $x$ to $x+N_r$;
\begin{align}
  \label{eq:prop}
  x &\xrightarrow{\omega_{r}(x)} x+N_{r},
\end{align}
where $N_{r} \in \mathbf{Z}^{D}$ is the transition step and is the
$r$th column in the \emph{stoichiometric matrix}
$N$. Eq.~\eqref{eq:prop} defines a continuous-time Markov chain over
the positive $D$-dimensional integer lattice.

When the reactions take place in a container of volume $\Omega$, it is
sometimes useful to know that the propensities often satisfy the
simple scaling law
\begin{align}
  \omega_{r}(x) &= \Omega u_{r}(x/\Omega)
\end{align}
for some function $u_{r}$ which does not involve $\Omega$. Intensities
of this form are called \emph{density dependent} and arise naturally
in a variety of situations \cite[Ch.~11]{Markovappr}.

\subsection{Mesoscopic diffusion}
\label{subsec:mesodiff}

\begin{figure}[htb!]
  \centering  
  \includegraphics[width=0.6\linewidth,height=0.6\linewidth]
    {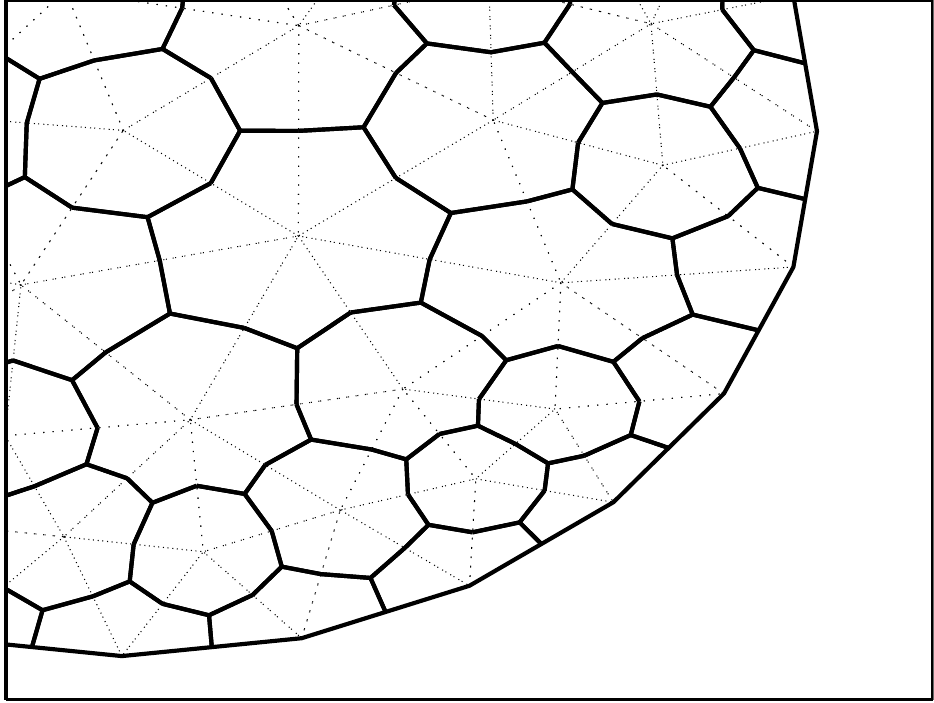}
  \caption{A 2D example of an unstructured triangular mesh. The primal
  mesh is shown in dashed and the dual in solid. Within each dual
  element the system is assumed to be well-stirred, and molecules can
  jump from each dual cell to the neighboring ones.}
  \label{fig:dual}
\end{figure}

In the mesoscale model, a diffusion event is modeled as a first order
reaction taking species $S_l$ in subvolume $\zeta_i$ from its present
subvolume to an adjacent subvolume $\zeta_j$,
\begin{equation}
  S_{li} \xrightarrow{a_{ij}\fatx_{li}} S_{lj},
\end{equation}        
where $\fatx_{li}$ is the number of molecules of species $l$ in
subvolume $i$. On a uniform Cartesian mesh such as those used in
MesoRD \cite{mesoRD}, the rate constant takes the value $a_{ij}=\gamma
/h^2$ where $h$ is the side length of the subvolumes and $\gamma$ is
the diffusion constant. In URDME we use an unstructured mesh made up
of tetrahedra and the rate constants are taken such that the expected
value of the number of molecules divided by the volume (the
concentration) converges to the solution obtained from a consistent
FEM discretization of the diffusion equation
\begin{equation}
  u_t = \gamma \Delta u.
\end{equation}
Using piecewise linear Lagrange elements and mass lumping, we obtain
the discrete problem
\begin{equation}
  \label{eq:FEM}
  u_t = M^{-1}Ku
\end{equation}
where $M$ is the lumped mass matrix and $K$ is the
stiffness matrix. The rate constants on the unstructured mesh are then
given by
\begin{equation}
  a_{ij} = \frac{1}{\Omega_i} k_{ij}, 
\end{equation}
where $\Omega_i$ is the diagonal entry of $M$ and can be interpreted
as the volume of the dual element associated with mesh node $i$ (see
Figure~\ref{fig:dual}). For more details, consult
\cite{SPDEPEFHL}.

The assumption made in the mesoscopic model is that molecules are
well-stirred within a dual cell. These dual cells correspond to the
cubes of the staggered grid in a Cartesian mesh.

\section{Algorithms}
\label{app:algorithms}

One of the most popular algorithms to generate realizations of the
CTMC in the well-stirred case is Gillespie's direct method (DM)
\cite{SSA}. Several algorithmic improvements of this method exist,
one of them being the next reaction method (NRM) due to Gibson and
Bruck \cite{NRM}.

The underlying algorithm in URDME is the next subvolume method (NSM)
\cite{BISTAB}. The NSM can be understood as a combination of NRM and
DM in order to tailor the algorithm to reaction-diffusion processes.

For reference, we first state below both DM and NRM and then outline
NSM.

\begin{algorithm}[htb!]
\caption{Gillespie's direct method (DM)}
\begin{algorithmic}
  \STATE{\textit{Initialize:} Set the initial state $\fatx$ and
  compute all propensities $\omega_r(\fatx), r=1, \ldots,
  \Mreactions$. Also set $t=0$.}

  \WHILE{$t < T$}

  \STATE{Compute the sum $\lambda$ of all the propensities.}

  \STATE{Sample the next reaction time (by inversion), $\tau = -
  \log(\rand)/\lambda$. Here and in what follows, `$\rand$'
  conveniently denotes a uniformly distributed random number in
  $(0,1)$ which is different for each occurrence.}
  
  \STATE{Sample the next reaction event (by inversion); find $n$ such
  that \newline $\sum_{j=1}^{n-1} \omega_j(\fatx) < \lambda \, \rand
  \le \sum_{j=1}^n \omega_j(\fatx)$}

  \STATE{Update the state vector, $\fatx = \fatx+N_n$ and set
  $t=t+\tau$.}

  \ENDWHILE

\end{algorithmic}
\label{alg:SSADM}
\end{algorithm}

\begin{algorithm}[htb!]
\caption{Gibson and Bruck's next reaction method (NRM)}
\begin{algorithmic}
  \STATE{\textit{Initialize:} Set $t = 0$ and assign the initial
  number of molecules. Generate the dependency graph $G$. Compute the
  propensities $\omega_{r}(\fatx)$ and generate the corresponding
  \emph{absolute} waiting times $\tau_{r}$ for all reactions
  $r$. Store those values in a heap $H$.}

  \WHILE{$t < T$}

  \STATE{Remove the smallest time $\tau_{n} = H_{0}$ from the top of
  $H$, execute the $n$th reaction $\fatx := \fatx+N_{n}$ and set $t :=
  \tau_{n}$.}

  \FORALL{edges $n \to j$ in $G$}

  \IF{$j \not = n$}

  \STATE{Recompute the propensity $\omega_{j}$ and update
    the corresponding waiting time according to
\begin{align*}
  \tau_{j}^{\mathrm{new}} &= t+\left(\tau_{j}^{\mathrm{old}}-t\right)
  \frac{\omega_{j}^{\mathrm{old}}}{\omega_{j}^{\mathrm{new}}}.
\end{align*}}

  \ELSE[$j = n$]

  \STATE{Recompute the propensity $\omega_{n}$ and generate a new
  absolute time $\tau_{n}^{\mathrm{new}}$. Adjust the contents of $H$
  by replacing the old value of $\tau_{n}$ with the new one.}

  \ENDIF

  \ENDFOR

  \ENDWHILE
\end{algorithmic}
\label{alg:NRM}
\end{algorithm}

\begin{algorithm}[htb!]
\caption{The next subvolume method (NSM)}
\label{alg:NSM}
\begin{algorithmic}
  \STATE{\textit{Initialize:} Compute the sum $\sigma_i^r$ of all
  reaction rates $\omega_{ri}$ and the sum $\sigma_i^d$ of all
  diffusion rates $a_{ij}\fatx_{si}$ in all subvolumes $i=1,\ldots
  ,\Ncells$. Compute the time until the next event in each subvolume,
  $\tau_i = -\log(\rand)/(\sigma_i^r+\sigma_i^d)$, and store all times
  in a heap $H$.}

  \WHILE{$t < T$}

  \STATE{Select the next subvolume $\zeta_n$ where an event takes
  place by extracting the minimum $\tau_n$ from the top of $H$.}
  
  \STATE{Set $t=\tau_n.$} 

  \STATE{Determine if the event in $\zeta_n$ is a reaction or a
  diffusion event. Let it be a reaction if $(\sigma_n^r+\sigma_n^d) \,
  \rand < \sigma_n^r$, otherwise it is a diffusion event.}

  \IF{Reaction event}
  
  \STATE{Determine the reaction channel that fires. This is done by
  inversion of the distribution for the next reaction given $\tau_n$
  in the same manner as in Gillespie's direct method in
  Algorithm~\ref{alg:SSADM}.}

  \STATE{Update the state matrix using the (sparse) stoichiometric
    matrix $N$.}

  \STATE{Update $\sigma_n^r$ and $\sigma_n^d$ using the dependency
  graph $G$ to recalculate only affected reaction- and diffusion
  rates.}

  \ELSE[Diffusion event]

  \STATE{Determine which species $S_{ln}$ diffuses and subsequently,
  determine to which neighboring subvolume $\zeta_{n'}$. This is
  again done by inversion using a linear search in the corresponding
  column of $D$.}

  \STATE{Update the state: $S_{nl}=S_{nl}-1$, $S_{n'l}=S_{n'l}+1$.}
  
  \STATE{Update the reaction- and diffusion rates of subvolumes
  $\zeta_n$ and $\zeta_{n'}$ using G.}
  
  \ENDIF
  
  \STATE{Compute a new waiting time $\tau_n$ by drawing a new random
  number and add it to the heap $H$.}
  
  \ENDWHILE 
\end{algorithmic}
\end{algorithm}

\end{document}